\documentclass[12pt,preprint]{aastex}

\newcommand{\kms}{\mbox{km s$^{-1}$}}
\newcommand{\etal}{\mbox{et al.}}

\newcommand{\nnh}{\mbox{N$_2$H$^+$ 1$\rightarrow$0}}
\newcommand{\Msun}{\mbox{$M_{\sun}$}}
\newcommand{\Lsun}{\mbox{$L_{\sun}$}}
\newcommand{\co}[2]{\mbox{CO $J = #1 \rightarrow #2$}} 
\newcommand{\jj}[2]{\mbox{$J = #1 \rightarrow #2$}}

\newcommand{\uco}[1]{\mbox{$^{#1}$CO}} 

\newcommand{\Tex}{\mbox{$T_{ex}$}}

\newcommand{\skipthis}[1]{}
\newcommand{\vlsr}{\mbox{$V_{lsr}$}}

\newcommand{\hh}{\mbox{H$_2$}}
\newcommand{\kp}{\mbox{K$'$}}
 
\slugcomment{to appear in ApJ Letters}
\shorttitle{BHR 71 Protobinary}
\shortauthors{Bourke}
  
\begin{document}

\title{IRAS 11590--6452 in BHR~71 -- a binary protostellar system?}
  
\author{Tyler L. Bourke}
\affil{Harvard-Smithsonian Center for Astrophysics, 60 Garden Street MS 42, 
Cambridge MA 02138, USA}
\email{tbourke@cfa.harvard.edu}

\begin{abstract}

New AAT near-infrared and SEST \uco{12}\ \jj{2}{1}\ observations are combined 
with existing ISO mid-infrared and ATCA cm radio continuum observations to
examine the protostellar content of the Bok globule BHR~71.  Together
with observations of Herbig-Haro objects, these data show:
(1) Two protostellar sources, IRS1 and IRS2, with a separation of 
$\sim$17\arcsec\ (3400 AU) are located within BHR~71.
(2) Each protostar is driving its own molecular outflow.  The outflow from 
IRS1 is much larger in extent, is more massive, and dominates the CO
emission.
(3) Both protostars are associated with Herbig-Haro objects and shock
excited 2.122 \micron\ \hh\ $v$=1-0S(1) emission, which coincide spatially
with their CO outflows.
(4) IRS1 is associated with cm continuum emission, with a flat or rising 
spectrum which is consistent with free-free emission, a
signpost of protostellar origin.

\end{abstract}



\section{Introduction}

BHR~71 (Bourke \etal\ 1995a,b) is a well isolated Bok globule located at
$\sim$200 pc, which harbors a highly collimated bipolar outflow 
(Bourke \etal\ 1997 -- hereafter B97).  The outflow is driven by a very
young Class 0 protostar with a luminosity of $\sim$9 \Lsun.

This paper brings together new observations with existing observations
to show that BHR~71 contains two embedded protostars with a
separation of $\sim$3400 AU.  Each protostar is driving a molecular
outflow seen in CO, but only one appears to be associated with a
substantial amount of circumstellar material.
The observations suggest that BHR~71 may contain an embedded binary 
protostellar system.


\section{Observations}

Near-infrared observations were undertaken with the 
128$\times$128 HgCdTe array camera IRIS on the 
Anglo-Australian Telescope (AAT).  Mounted at
the f/36 Cassegrain focus IRIS provided a field of view of approximately
100\arcsec\ with a resolution of 0\farcs79 pixel$^{-1}$.  A 4$\times$4
mosiac at K$'$ (2.11 \micron) with 90\arcsec\ offsets between frames was
obtained on 1992 February 15, with an integration time of 200s/frame.  A
3$\times$3 mosaic in the \hh\ $v$=1-0S(1) transition 
at 2.12 \micron\ (1\% bandpass) with 80\arcsec\ offsets 
was obtained on 1993 January 11, with an integration time of 60s/frame.
Standard data reduction was performed with the Starlink FIGARO data reduction
package.

Observations of the \uco{12}\ \jj{2}{1}\ transition at 230537.99 MHz 
were obtained with the 15-m Swedish ESO Submillimetre Telescope (SEST) 
during 2000 May.  The backend used was an
acousto-optical spectrometer providing a channel separation of 43 kHz
(0.055 \kms) over 2000 channels.  The observations were performed in
dual beam-switching mode with a beam separation of 11\arcmin47\arcsec\
in azimuth.  Systems temperatures of 220~K were recorded during the 
observations.  A small 13 point map with 30\arcsec\
offsets about the position of the embedded source BHR~71-mm (B97) was made, 
with an integration time of 60s/point.  
The beamsize at this frequency is $\sim$23\arcsec.

BHR 71 was observed by the Infrared Space Observatory (ISO) on 1996
August 19 as part of program DMARDONE.  Observations 
in the LW2 (5.0--8.5 \micron) and LW3 (12.0--18.0 \micron) bands were
obtained with a field of view of $\sim$90\arcsec.  Full details of the
observations can be found in van den Ancker, Mardones \& Myers (2001;
see also Myers \& Mardones 1998).

The Australia Telescope Compact Array (ATCA)\footnote{
The ATCA is part of the Australia Telescope which is funded by the 
Commonwealth of Australia for operation as a National Facility managed
by CSIRO.} was used to obtain images
of BHR 71 at 3 and 6 cm (8.64 and 4.80 GHz respectively) as part of
program C368 (Wilner \etal\ 2001).  Observations were made on 1994
November 12 with the 6D configuration, observing both frequencies
simultaneously with 128 MHz bandpasses.  The data was reduced with
MIRIAD and imaged using natural weighting, resulting in beamsizes of 
$\sim$2\arcsec\ at 3 cm and $\sim$4\arcsec\ at 6 cm.  Full details can be 
found in Wilner \etal\ 2001.


\section{Results}

The near-infrared (NIR) images are shown in Figure~\ref{fig-nir}.  The \kp\
image is shown in (a), and the narrowband 2.12 \micron\ image in (b).
No continuum subtraction has been performed on the narrowband image.
It is immediately evident by comparison of the two images that most of the 
non-stellar emission is due to the emission in the \hh\ $v$=1-0S(1) line.  
This is most likely due to shocks in the outflowing gas (Eisl\"{o}ffel
1997).  In BHR~71
the large scale CO outflow lies at a PA of $\sim$165\degr\ (B97) and so is 
well aligned with the near-infrared emission.

Mid-infrared (MIR) emission in the ISO LW2 band is overlayed on the \kp\ 
image in Fig.~\ref{fig-nir}(a), labelled as ``ISO 7\micron".  Two of the
7 \micron\ sources appear to be located at the apexes of NIR emission,
strongly suggesting that they are associated with the emission.  Source
``1" (hereafter IRS1) lies at the apex of the reflection nebulosity seen
also in the I band image presented by B97, which is
associated with the large blue-shifted CO outflow lobe.  IRS1 is also
co-incident with the position of the mm source BHR~71-mm, also known as
IRAS 11590-6452 (B97).  No such counterpart exists for source ``2" (hereafter
IRS2), which is the weaker of the two sources at 7 \micron.  The fluxes
for IRS1 and IRS2 in the ISO LW2 (7 \micron) and LW3 (15 \micron) bands
are listed in Table~\ref{tbl-pos}.  IRS2 is not seen directly at 2
\micron.  The NIR feature coincident with IRS2 in Fig.~\ref{fig-nir}(a) is
non-stellar, by comparison of its PSF with stars in the same image.

One centimeter continuum source was detected toward BHR~71, at both 3 and
6~cm (Wilner \etal\ 2001).  The position of the source at 3~cm is indicated in 
Fig.~\ref{fig-nir}(b) and is listed in the notes to Table~\ref{tbl-pos}.  
The free-free emission coincides with the
position of IRS1 and has a spectral index at centimeter wavelengths
that is consistent with a flat or rising spectrum due to free-free emission, a
signpost of protostellar origin (Rodr\'{\i}guez 1994).  

Corporon \& Reipurth (1997) discovered two Herbig-Haro associations in
BHR~71 -- HH~320 and HH~321.  Their locations are shown on
Fig.~\ref{fig-nir}(b).  The positions of the HH objects as listed in
Corporon \& Reipurth (1997) are incorrect (P. Corporon, private
communication).  As discussed below, the plate solutions for their
\ion{S}{2}\ image have been redetermined and the positions of HH~320 and
HH~321 remeasured.  The correct positions, accurate to $<$ 1\arcsec, are
given in Table~\ref{tbl-pos}.  From Fig.~\ref{fig-nir}(b) it can be seen that
HH~320 is coincident with the NIR emission associated with IRS2, while
HH~321 is coincident with the NIR emission associated with IRS1.

In Figure~\ref{fig-outflow}(a) is shown an enlarged view of the central
part of the 2.12 \micron\ image presented in Fig.~\ref{fig-nir}(b).
Indicated on this figure are the locations of the ISO sources IRS1 and
IRS2 (unfilled squares).  Overlayed on the figure are contours of
integrated \uco{12}\ \jj{2}{1}\ emission, with solid contours
representing emission that is blue-shifted ($-20 < \vlsr < -6$ \kms)
with-respect-to the cloud systemic velocity (\vlsr\ $\sim -4.5$ \kms;
B97), and dotted contours representing red-shifted emission
($-3 < \vlsr < 30$ \kms).  The large scale CO outflow mapped
by B97 is evident as open blue- and red-shifted
contours, orientated approximately north-south with IRS1 lying between
them.  On
scales larger than is mapped in the present \co{2}{1}\ observations this
outflow has a position angle of $\sim$165\degr.  Closed contours of
blue-shifted emission peak at the position of HH~320.

In Figure~\ref{fig-outflow}(b) is shown the CO \jj{2}{1}\ spectrum at the
($-30$\arcsec,30\arcsec) offset position (HH~320).  The
high velocity outflow wing emission at this position due to IRS1
(red-shifted) and IRS2 (blue-shifted) is clearly seen.

In order to determine accurately the ISO positions and the relative
positions of the centimeter source and HH objects it was necessary to
register the images to the same spatial system.  First, a plate solution
for a 40\arcmin\ FOV Digital Sky Survey (DSS)
image centered on BHR~71 was determined by comparison with 25 stars from 
the {\em HST} Guide Star Catalog (GSC).
Using both the corrected DSS image and the GSC, 
plate solutions for the I band image and the \ion{S}{2}\ image (Corporon \&
Reipurth 1997 -- kindly provided in digital form by P.\ Corporon) were 
determined, using 14 stars common to all three bands.  The
uncertainty in positions measured from these frames is believed to be
$<$1\arcsec, by comparison with the GSC and with each other.

The plate solution for the NIR images was determined using stars common with
the I band image, 20 for the \kp\ image and 10 for the narrowband 2.12
\micron\ image, respectively.  Finally, the ISO 7 \micron\ plate solution was
determined using the three ISO sources visible in Fig.~\ref{fig-nir}(a)
and not associated with NIR nebulosity.  The positions of the HH objects
were measured directly from the corrected \ion{S}{2}\ image, using the
peak positions due to their non-symmetric shape.  All positions quoted
here are believed to be accurate to $\leq$1\arcsec.


\section{Discussion}

\subsection{Two protostars -- two outflows}

The data presented here clearly indicates that two protostellar sources
are present in BHR~71, each driving its own molecular outflow.
IRS1 and its large-scale outflow has been discussed in detail by 
B97.  The spectacular shock chemistry in the outflow
has been studied by Garay \etal\ (1998).  B97 discovered
extended 1.3 mm emission associated with IRS1 and suggested that this is
due to a massive circumstellar disk.  

IRS2 has been noted previously (Myers \& Mardones 1998).  The data
presented here shows that it is driving a compact CO outflow and is
associated with near-infrared molecular hydrogen emission 
(Fig.~\ref{fig-outflow}(a)), which is coincident with HH~320 and most likely 
shock excited (Eisl\"{o}ffel 1997).  
Earlier evidence for the blue-lobe can be seen in the channel maps 
of the IRS1 outflow presented in B97.  Their Figure 6
shows the emission from the CO \jj{1}{0}\ transition, and the outflow
associated with IRS2 can be seen in the velocity range
$-10.8 <$ \vlsr\ $<-8.4$ \kms.   

Comparison of the I band image and the IRS1 outflow suggests that its blue 
lobe has broken through the near side of the globule (B97).
The presence of HH objects in the blue lobe of the IRS2 outflow suggests
that this outflow is also penetrating the near side of the globule.  
The relative mass of the blue and red outflow lobes of IRS1 suggests
that the red lobe of the IRS1 outflow has not penetrated the far side
of the globule (B97).  It is likely that IRS1 and IRS2
are located at a similar depth within the globule from the near side,
and that their angular separation of $\sim$17\arcsec\ is a good indication of 
their physical separation, which is therefore 3400 AU for the assumed
distance of 200 pc.  

The limited CO observations do not reveal a red-shifted lobe to the IRS2 
outflow.  Due to the large inclination of the IRS1 outflow to the line-of-sight,
both blue- and red-shifted outflow emission is seen in the SE part of
the IRS1 outflow (B97), which probably masks any emission from the red-shifted
lobe of the IRS2 outflow.  However, the NIR image 
shows a knot of \hh\ line emission, arrowed in Fig.~\ref{fig-nir}(b),
which is a possible counterflow to the IRS2 blue-shifted outflow lobe.
A line drawn from this knot through and extending past IRS2 bisects the
limb-brightened conical reflection nebulosity (connecting IRS2 and HH~320)
at the base of the blue-outflow lobe of IRS2.  Taking this to define the 
outflow axis, the IRS2 outflow has a P.A. of $\sim-36$\degr.
 
Combining the IRS2 CO \jj{2}{1}\ data with the CO \jj{1}{0}\
data of B97 for IRS2 allows for an estimation of the CO excitation 
temperature \Tex, though
the signal-to-noise of the latter data set is poorer, and the sampling
grids are different.  
As no isotopic data are available, optically thin conditions are
assumed, which implies \Tex\ $\sim$10 K.  The mass of the blue-shifted
emission in the IRS2 outflow is $\sim$0.004 \Msun, significantly less
than the $\sim$0.06 \Msun\ in the blue-shifted gas of the SE lobe of
the IRS1 outflow (B97 - using CO
\jj{1}{0}\ data and assuming optically thin conditions).
The IRS1 outflow is moderately optically thick ($\tau \sim 2$; B97), so the 
optical depth of the IRS2 outflow would have to be significantly greater 
to modify this comparison.
Although significantly less massive than the IRS1
outflow, the IRS2 outflow is still able to produce shock-excited \hh\
emission and HH objects.

Previous observations by B97 show that IRS1 is a Class 0 protostar.
Compared to IRS1, IRS2 is much weaker at 7 \micron\ and drives a much
less massive outflow, and is not detected at 1.3~mm (\S~\ref{sec-binary}).
This suggests that IRS2 is more evolved than IRS1 and is most likely a
Class I protostar.

\subsection{A binary protostellar system?}
\label{sec-binary}

The presence of two sources in BHR~71 with a separation of 3400 AU, neither 
detected at wavelengths $<$ 7 \micron, and both driving molecular outflows,
suggests that a binary protostellar system has formed within the globule.
Observations by B97 show that strong 1.3 mm continuum
emission is associated with the protostellar pair, but the emission is highly
peaked on IRS1, and no obvious extension including IRS2 is seen (their
Figs. 9 \& 13).  
The lack of mm emission and the weakness of the 7 \micron\ emission
compared to IRS1 suggests that IRS2 is not surrounded by a significant
amount of circumstellar dust.
This situation is similar to that observed in Bok globule CB230 
(Launhardt 2001).  High-angular resolution observations ($\sim$2\arcsec) of 
CB230 show compact mm emission associated with only one component of a NIR
protostellar pair separated by $\sim$10\arcsec, suggesting that like
BHR~71 only this component has a substantial circumstellar disk, though 
both protostars in CB230 drive CO outflows.  Another wide binary
protostellar pair, SVS~13 (Bachiller et al.\ 1998; 
separation 4300 AU), shows mm emission from both protostars.

Kinematic information indicating a common center of gravity is required
to show that a stellar pair is a binary.
Launhardt (2001) has also observed \nnh\ emission (93.7 GHz) from CB230
with a resolution of $<$10\arcsec\ and found two cores with a separation
of 10\arcsec\ (4500 AU), each spatially
coincident with a NIR source, and which rotate about the axis
perpendicular to the axis joining the cores.   This implies that CB230
contains a true binary protostellar system.
Molecular line observations to date of BHR~71 (B97) 
do not have sufficient angular resolution to determine if IRS1 and IRS2 
are each associated with their own molecular core.   
Though there are a similarities between BHR~71 and CB230 -- both contain
2 protostellar sources in a Bok globule, each driving a CO outflow, and 
large scale mm emission is centered on only one of the sources -- BHR~71 can 
only be considered as a candidate binary protostellar system until high-angular
resolution molecular line observations become available.


\section{Conclusions}

New near-infrared and \uco{12}\ \jj{2}{1}\ observations have been combined 
with existing ISO mid-infrared and ATCA cm radio continuum observations to
examine the protostellar content of the Bok globule BHR~71.  Together
with observations of Herbig-Haro objects, these data show:

(1) Two protostellar sources, IRS1 and IRS2, with a separation of 
$\sim$17\arcsec\ (3400 AU) are located within BHR~71, as revealed by ISO
mid-infrared observations.  IRS1 is the brighter of the two by about a
factor 10. 

(2) Each protostar is driving its own molecular outflow, as revealed by
the \uco{12}\ \jj{2}{1}\ observations.  The outflow from IRS1 is much
larger in extent, is more massive, and dominates the CO emission.

(3) Both protostars are associated with Herbig-Haro objects and shock
excited 2.122 \micron\ \hh\ $v$=1-0S(1) emission, which coincide spatially
with their CO outflows.

(4) IRS1 is associated with cm continuum emission, with a flat or rising 
spectrum which is consistent with free-free emission, a
signpost of protostellar origin.

The observations suggest that a binary protostellar system has formed within
BHR~71.
IRS1 is a Class 0 protostar, while IRS2, associated with much less
circumstellar dust and driving a much weaker CO outflow, is probably a
more evolved Class I protostar.
High angular resolution mm molecular line observations are required to
determine if IRS1 and IRS2 are a physically bound binary.

\acknowledgements

I thank David Wilner, Diego Mardones and Mario van den Ancker for
sharing data in advance of publication.  I also thank Patrice Corporon
for providing his \ion{S}{2}\ image in digital form.   I am grateful 
to Garry Robinson and A.R. Hyland for fruitful collaborations.


\begin{deluxetable}{lcccc}
\tablecolumns{5}
\tablewidth{250pt}
\tablecaption{Positions and Fluxes \label{tbl-pos}}
\tablehead{
\colhead{} & \multicolumn{2}{c}{Position (B1950)} & \colhead{}
& \colhead{} \\
\cline{2-3}
\colhead{} & \colhead{R.A.} & \colhead{Dec.} &
\colhead{7 \micron} & \colhead{15 \micron} \\ 
\colhead{Name}        & \colhead{$h$~~$m$~~$s$\phd\phd\phd}      &
\colhead{\phs\arcdeg~~~\arcmin~~~\arcsec} &  \colhead{(mJy)} &
\colhead{(mJy)}  
}
\startdata
\sidehead{ISO sources}
IRS1\tablenotemark{(a)} & 11~59~02.7 & $-$64~52~06 & 346 & 230 \\
IRS2 & 11~59~00.1 & $-$64~52~02 & \phn39 & $<$18 \\
\sidehead{Herbig-Haro objects\tablenotemark{(b)}}
HH~320 A & 11~58~58.3 & $-$64~51~31 & \nodata & \nodata \\
\phm{HH~320} B & 11~58~57.4 & $-$64~51~23 & \nodata & \nodata \\
HH~321 A & 11~59~02.5 & $-$64~52~51 & \nodata & \nodata \\
\phm{HH~321} B & 11~59~05.2 & $-$64~53~35 & \nodata & \nodata 
\enddata
\tablenotetext{(a)}{The 3 cm position of IRS1 is R.A.\ (B1950) = $11^{\rm
h}59^{\rm m}02\fs61$ and Dec.\ (B1950) = $-64\arcdeg52\arcmin06\fs7$.}
\tablenotetext{(b)}{Positions are different to those given in Corporon \&
Reipurth (1997) -- see text.}
\end{deluxetable}

\clearpage

\figcaption[f1.eps]
{(a) -- \kp\ image of BHR~71 (greyscale) overlayed with ISO LW2
contours (5.0--8.5 \micron).  The approximate region observed by ISO is 
indicated by
the dashed box, and the embedded protostars IRS1 (``1'') and IRS2 (``2'')
are labeled.  (b) -- Narrowband 2.12 \micron\ + continuum image
(greyscale).  The positions of HH~320 and HH~321 (Corporon \& Reipurth
1997) are marked with crosses, and the position of the 3 cm continuum source is
marked with an unfilled box.
\label{fig-nir}}

\figcaption[f2.eps]
{(a) -- Narrowband 2.12 \micron\ + continuum image (greyscale) 
overlayed with
contours of \co{2}{1}\ emission.  The blue (red) contours represent
emission integrated over the velocity range $-20 < \vlsr < -6$ \kms\ 
($-3 < \vlsr < 30$ \kms), which is 
blueshifted (redshifted) with-respect-to the cloud systemic velocity.  
The positions of IRS1 and IRS2 are marked with unfilled boxes.  Offsets
are relative to R.A.\ (B1950) = $11^{\rm
h}59^{\rm m}02\fs3$ and Dec.\ (B1950) = $-64\arcdeg52\arcmin01\arcsec$.
(b) -- CO \jj{2}{1}\ spectrum at the ($-30$\arcsec,30\arcsec) offset 
position.  The dashed lines represent the limits of the
ambient cloud emission ($-6 < \vlsr < -3$ \kms).  High velocity blue
(red) shifted line wings are present at velocities $\vlsr < -6$ \kms\
($\vlsr > -3$ \kms).
\label{fig-outflow}}

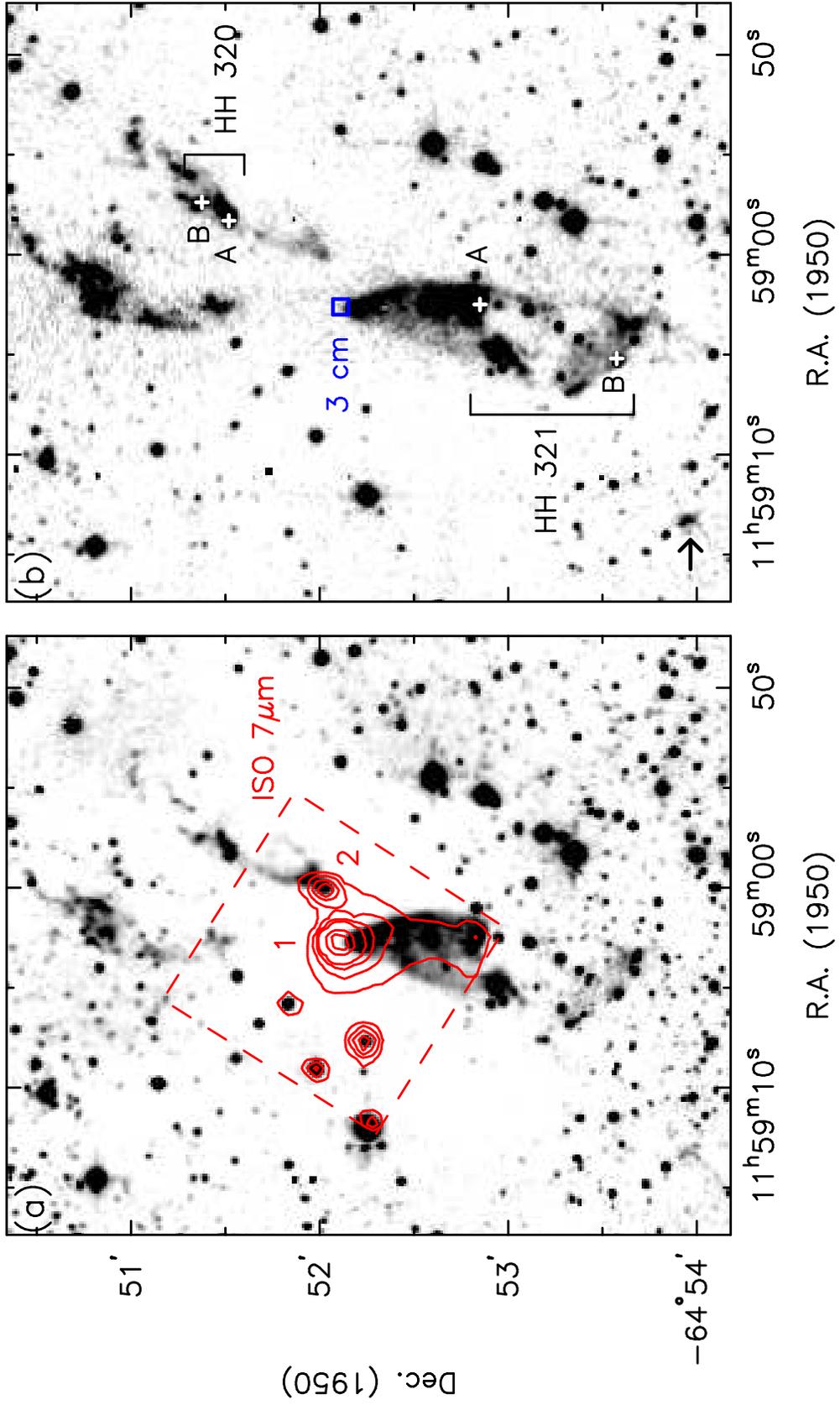
\begin{figure}
\figurenum{1}
\epsscale{0.8}
\plotone{f1.cps}
\caption{}
\end{figure}
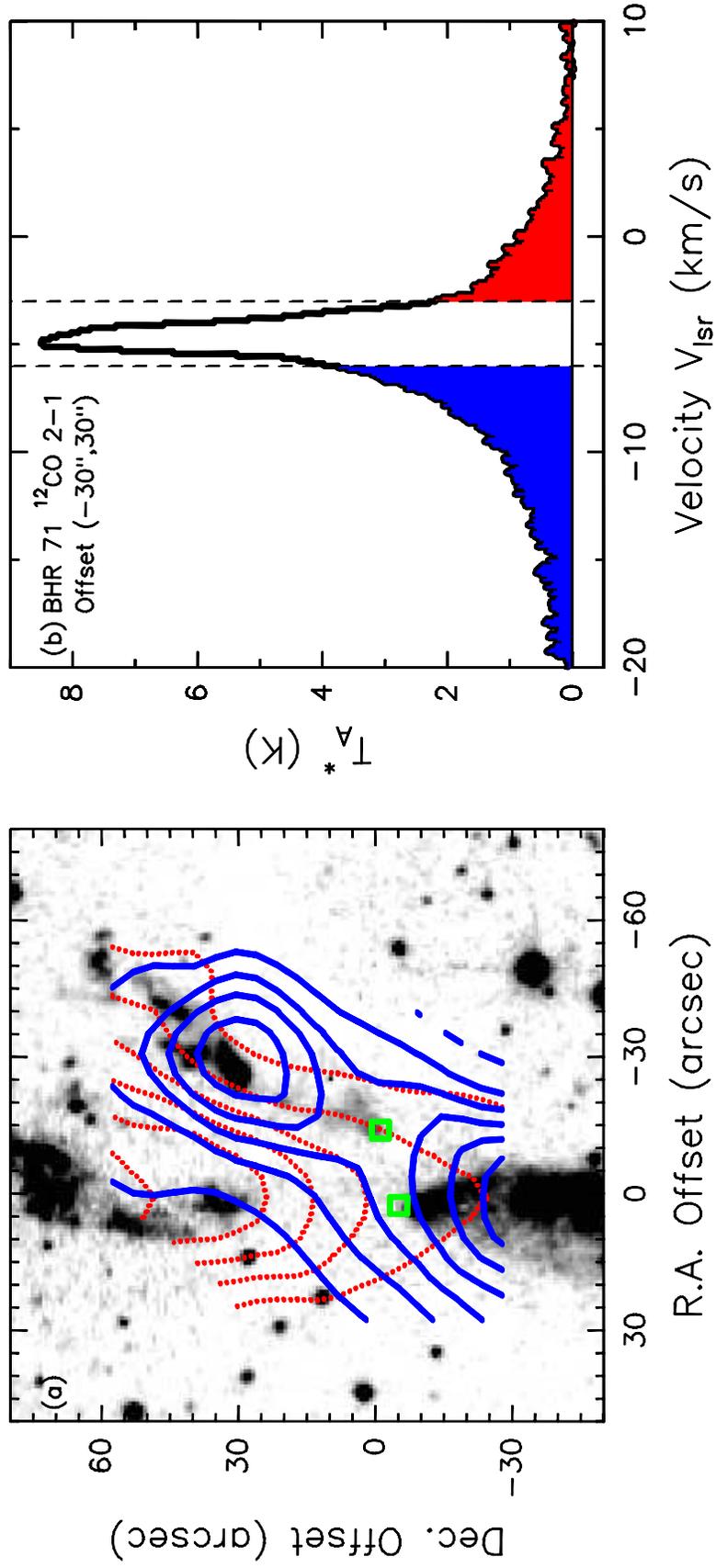
\begin{figure}
\figurenum{2}
\epsscale{0.7}
\plotone{f2.cps}
\caption{}
\end{figure}

\end{document}